\documentclass[epj]{webofc}

\newcommand{\Z}{\ensuremath{\mathcal{Z}}}
\newcommand{\M}{\ensuremath{\mathcal{M}}}
\newcommand{\K}{\ensuremath{\mathcal{K}}}

\newcommand{\diff}[1]{\mbox{d}{#1} \, }

\usepackage[utf8]{inputenc}
\usepackage[varg]{txfonts}   
\usepackage{booktabs}
\usepackage{xcolor}
\definecolor{darkred}{rgb}{0.4,0.0,0.0}
\definecolor{darkgreen}{rgb}{0.0,0.4,0.0}
\definecolor{darkblue}{rgb}{0.0,0.0,0.4}
\definecolor{fzjblue}{HTML}{005B81}
\usepackage[bookmarks,linktocpage,colorlinks,
    linkcolor = darkred,
    urlcolor  = darkblue,
    citecolor = darkgreen]{hyperref}

\usepackage{feynmp-auto}    
\usepackage{simplewick}     
\usepackage{braket}         
\usepackage{tensor}         
\usepackage{bbm}            
\usepackage{mathrsfs}       

\usepackage[colorinlistoftodos,prependcaption,textsize=tiny]{todonotes}
\usepackage{subcaption}
\usepackage{tikz}
\usetikzlibrary{positioning, calc}
\graphicspath{{./figures/}}

\wocname{EPJ Web of Conferences}
\woctitle{Lattice2017}
\begin{document}
%
\selectlanguage{english}
\title{%
Hubbard-Stratonovich-like Transformations for Few-Body Interactions
}
\author{%
\firstname{Christopher} \lastname{K\"orber} \inst{1} \and
\firstname{Evan}        \lastname{Berkowitz}\inst{1} \and
\firstname{Thomas}      \lastname{Luu}      \inst{1}
}
\institute{%
  Institut f\"ur Kernphysik and
  Institute for Advanced Simulation   \\
  Forschungszentrum J\"ulich          \\
  52425 J\"ulich Germany
}
\abstract{%
Through the development of many-body methodology and algorithms, it has become possible to describe quantum systems composed of a large number of particles with great accuracy.
Essential to all these methods is the application of auxiliary fields via the Hubbard-Stratonovich transformation.
This transformation effectively reduces two-body interactions to interactions of one particle with the auxiliary field, thereby improving the computational scaling of the respective algorithms.
The relevance of collective phenomena and interactions grows with the number of particles.
For many theories, e.g. Chiral Perturbation Theory, the inclusion of three-body forces has become essential in order to further increase the accuracy on the many-body level.
In this proceeding, the analytical framework for establishing a Hubbard-Stratonovich-like transformation, which allows for the systematic and controlled inclusion of contact three- and more-body interactions, is presented.
}
\maketitle


\section{Introduction}\label{sec:introduction}
The Hubbard-Stratonovitch (HS)  transformation \cite{Stratonovich1957,Hubbard1959} is a common tool in many areas of theoretical physics, ranging from condensed matter to nuclear physics \cite{gubernatis2016quantum,Kleinert:2011rb}.
Though the basic idea---completing the square---is quite simple, the practical benefits can be numerous.
It is used to linearize the Hamiltonian in terms of its density or number operator by the introduction of an auxiliary field, which eventually is integrated out.
Thus, a many-body system can effectively be described as a one-body system which interacts with a background field, drastically simplifying the numerical description of many-body problems \cite{Lee:2008fa,Borasoy2007,Epelbaum:2009,Carlson:2014vla}.

In this proceeding, which closely follows \cite{Korber:2017emn}, we generalize the HS transformation to linearize arbitrary many-body systems by the introduction of just one auxiliary field, which is distributed according to specific probability distributions that allow the inclusion of different auxiliary field interactions.
The presentation is organized as follows: in section \ref{sec:the-basic-idea} we review the basic idea of the original HS transformation from a more diagrammatic perspective and explore how it could possibly be generalized.
In section \ref{sec:the-formalism} we discuss the formalism which enables the identification of a linearized Hamiltonian including auxiliary fields which eventually corresponds to a specific many-body Hamiltonian after integrating out the auxiliary fields.
A case study for an analytically solvable two-site model is presented in section \ref{sec:numerics} and we finally discuss our results in section \ref{sec:discussion}.

\section{The Basic Idea}\label{sec:the-basic-idea}
In the original Hubbard-Stratonovich transformation one linearizes the action by the introduction of a background field $\phi \in \mathbbm{R}$, which couples to the usual fermionic (or bosonic) fields $\psi$.
From now on, we denote the fermionic (or bosonic) bilinear by the density operator $ \hat \rho \equiv \bar \psi \psi $.
After integrating over this auxiliary field, one recovers the original interaction
\begin{equation}\label{eq:def:HS}
		\exp\left\{ + \lambda^{(2)} \hat \rho^2 \right\}
		=
	\frac{1}{\sqrt{\pi}}
	\int \mathrm{D}{\phi}
	\exp\left\{ - \phi^2 -  2 \sqrt{ \lambda ^{(2)} } \, \phi \hat \rho  \right\}
	\, .
\end{equation}
From a diagrammatical point of view, the auxiliary field can be interpreted as a bosonic spin-zero field with a Yukawa-like fermion interaction (and a constant propagator).

\begin{figure}[htb]
  \centering
  \begin{tikzpicture}[node distance=5mm]
    \node               (a) {\includegraphics[width=0.2\textwidth]{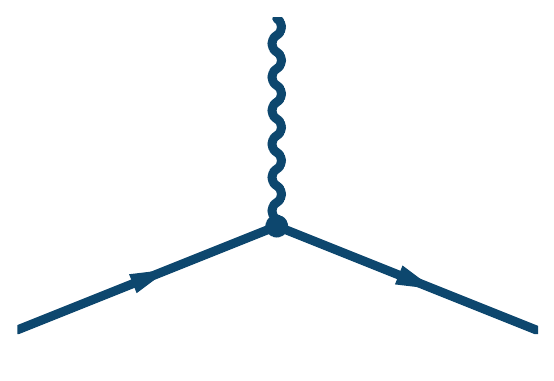}}  node[below=-5mm of a] {(a)\label{fig:HS-diagramatic-a}};
    \node[right = of a] (b) {\includegraphics[width=0.2\textwidth]{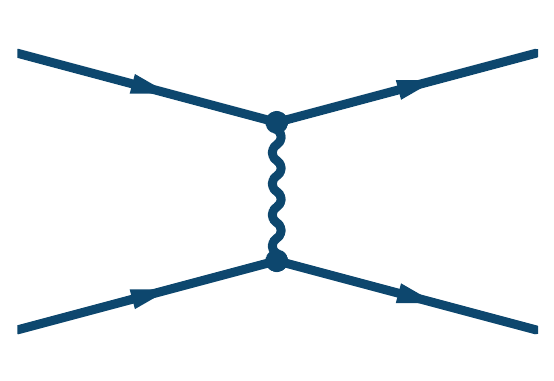}} node[below=-5mm of b] {(b)};
    \node[right = of b] (c) {\includegraphics[width=0.2\textwidth]{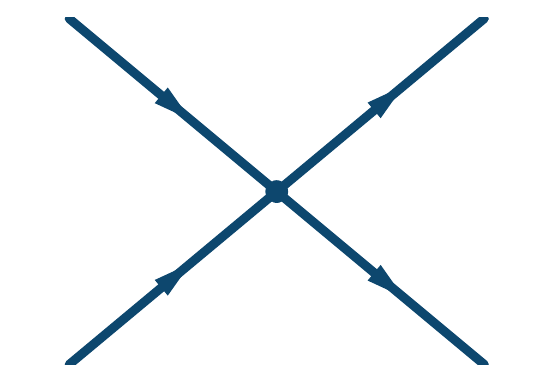}}    node[below=-5mm of c] {(c)};
    \draw [->, line width=0.5mm, draw=fzjblue] (a) -- (b);
    \draw [->, line width=0.5mm, draw=fzjblue] (b) -- (c);
  \end{tikzpicture}
  \caption{\label{fig:HS-diagramatic}Diagramatic interpretation of the original Hubbard-Stratonovich transformation.  One computes all possible diagrams allowed by the interactions [from (a) to (b)] and integrates out the auxiliary field to obtain an effective interaction [from (b) to (c)].}
\end{figure}

If one interprets the right-hand-site of equation (\ref{eq:def:HS}) as a path integral, the diagrammatic interpretation would be the sum of products of two-fermion diagrams which exchange an auxiliary field.
Integrating out the auxiliary field, which diagrammatically corresponds to contracting the auxiliary field lines, effectively generates the desired two-body fermion interaction (see figure \ref{fig:HS-diagramatic}).

The generalization of this diagrammatic interpretation is the basic idea to include three-body forces.
We start with a possible diagram which consists of one-body fermion and any possible $N$-body auxiliary field interaction.
\begin{figure}[htb]
  \centering
  \begin{subfigure}[b]{0.17\textwidth}
    \centering
    \includegraphics[width=\textwidth]{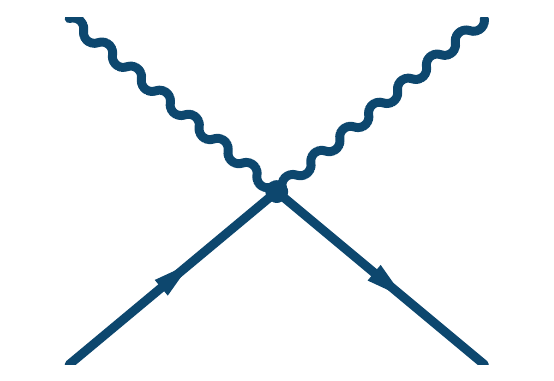}
    \caption{\label{fig:F2A2}}
  \end{subfigure}
  \quad
   \begin{subfigure}[b]{0.17\textwidth}
    \centering
    \includegraphics[width=\textwidth]{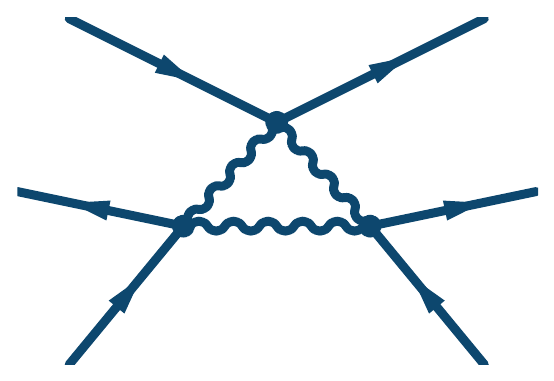}
    \caption{\label{fig:3F2A2}}
  \end{subfigure}
  \quad
  \begin{subfigure}[b]{0.17\textwidth}
    \centering
    \includegraphics[width=\textwidth]{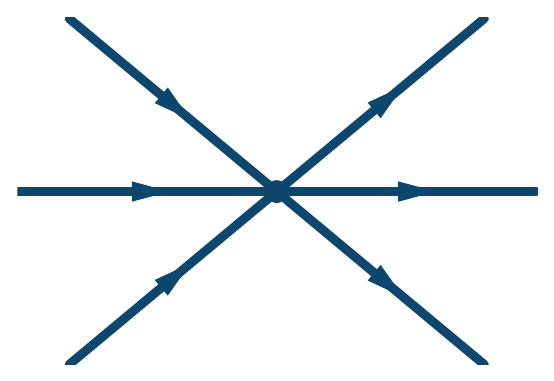}
    \caption{\label{fig:3F}}
  \end{subfigure}
  \quad
  \begin{subfigure}[b]{0.17\textwidth}
    \centering
    \includegraphics[width=\textwidth]{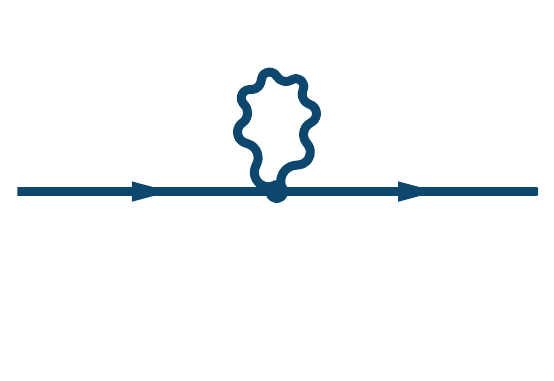}
    \caption{\label{fig:1F}}
  \end{subfigure}
  \begin{subfigure}[b]{0.17\textwidth}
    \centering
    \includegraphics[width=\textwidth]{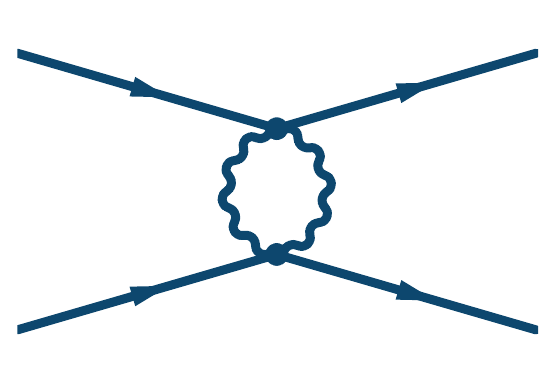}
    \caption{\label{fig:2F}}
  \end{subfigure}
  \caption{\label{fig:Three-body-HS-diagramatic}Allowed diagrams and  induced forces due to transformation with a quadratic auxiliary field, one-body fermion interaction.}
\end{figure}

The simplest example is the case of a one-body fermion, two-body auxiliary field interaction (figure \ref{fig:F2A2}) which can generate a three-body fermion interaction (figure \ref{fig:3F2A2}).
In order to guarantee a convergent integrand, the highest power of auxiliary fields appearing in the exponential should be at least $\phi^4$.
After contracting the auxiliary field lines, one obtains the desired three-fermion force (figure~\ref{fig:3F}).
However, when one actually computes all possible diagrams, one generates an infinite tower of different $N$-fermion diagrams (see figures \ref{fig:1F}, \ref{fig:2F} ...).
The only way to include three- and higher-body forces in a controlled fashion is to identify the  induced $N$-fermion couplings, depending on the auxiliary field interaction coefficients, and, at best, find a convergent and controllable expansion
\begin{equation}\label{eq:Zla}
	\int \mathrm{D}{\phi}
	\exp\left\{ - \phi^4 -  c_2 \, \phi^2 \hat \rho  \right\}
	=
	\sum_{M=0}^\infty \mathcal{Z}^{(M)}_{c, 4} \, \hat \rho^M
	\overset{!}{=} 
	\exp\left\{
		-\sum\limits_{k=1}^{\infty} \lambda^{(k)} \hat \rho ^k
	\right\}
	\equiv
	\Z_{\lambda}
	\, .
\end{equation}

\section{The Formalism}\label{sec:the-formalism}
In order to stay as general as possible, we do not limit ourselves to quadratic interactions between the auxiliary field and the fermion bilinears.
The most general form with one type of auxiliary-field self-interaction is given by the following integral
\begin{align}\label{eq:Zc}
	\Z_{c, N} 
	&\equiv
	\int\limits_{-\infty}^\infty \diff{\phi} P_N(\phi)
		\exp\left\{ - \sum\limits_{j=1}^{2N-1} c_j \phi^j \hat \rho \right\}\, , &
	P_N(\phi) &= 	\frac{N}{ \Gamma \left( \frac{1}{2N}  \right) } e^{ - \phi^{2N}} \, .
\end{align}
where $\hat \rho = \sum_{f} {\bar{\psi}}_{f} \psi_{f}$ is the fermion density operator, $f$ runs over the different fermion species at a given site and $P_N(\phi)$ is the normalized probability distribution.
This integral is well defined for any coefficients $c_i \in \mathbbm C$ and $N \in \mathbbm N$.
It is sufficient to consider auxiliary fields at just one point in space-time, because the the density operators at different sites commute and the auxiliary fields get no kinetic term.

The argument of the exponential in \eqref{eq:Zc} describes all possibly allowed interactions between the one-fermion operator $\hat \rho$ and auxiliary fields, with $j=1 \ldots 2 N -1$ open auxiliary field lines of strength $c_j$.
Accordingly, the integral is normalized in the case of $c_j = 0$ $\forall j \Rightarrow \Z_{c, N} = 1$, which corresponds to the absence of auxiliary fields.
For any $c_j \neq 0$, one effectively induces $N$-body forces after integrating out the auxiliary fields.
Hence, to find a controlled description of $N$-body fermion forces $\lambda^{(k)}$ in terms of auxiliary field coefficients $c_j$, one has to match both expressions order by order in the operator expansion of the exponential
\begin{align}
	\Z_{c, N}
	&=
	\sum_{M=0}^\infty
	\Z_{c, N}^{(M)} \hat \rho^M \, , &
	\Z_{\lambda}
	&=
	\sum_{M=0}^\infty
	\Z_{\lambda}^{(M)} \hat \rho^M \, , &
	\Z_{c, N}^{(M)} & \overset{!}{=} \Z_{\lambda}^{(M)} \quad \forall M \, .
\end{align}

\begin{figure}[htb]
  \centering
  \begin{tikzpicture}[node distance=4mm]
    \node                (11) {\includegraphics[width=0.15\textwidth]{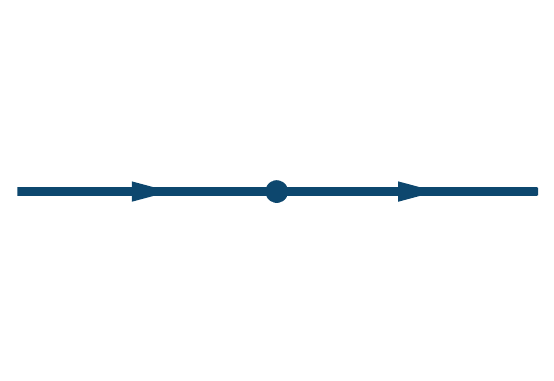}}         node[above=-8mm of 11] {$\lambda^{(1)}$}; 
    \node[right = of 11] (12) {\includegraphics[width=0.15\textwidth]{F2A2_A2_L1}} node at ($(11)!0.5!(12)$) {$=$};
    \node[right = of 12] (13) {\includegraphics[width=0.15\textwidth]{F2A1}}       node[below=-7mm of 13] {$c_1$};
    \node[right = of 13] (14) {\includegraphics[width=0.15\textwidth]{F2A2}}       node[below=-7mm of 14] {$c_2$};
    \node[right = of 14] (15) {\includegraphics[width=0.15\textwidth]{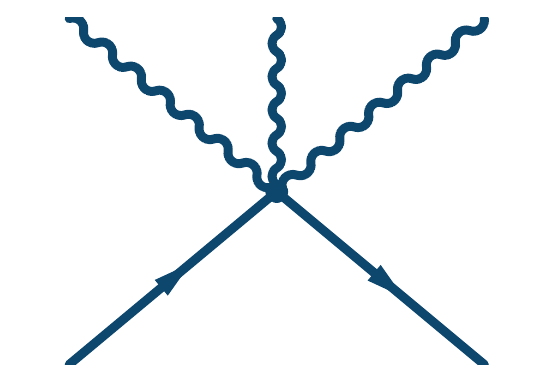}}       node[below=-7mm of 15] {$c_3$};
    \draw[thick, fzjblue] ($(13.north west)+(-0.15,0.15)$) rectangle ($(15.south east)+(0.15,-0.15)$);
    \node[below=-7mm of 12] {\small $c_2 \times \begin{cases}\gamma\\\frac{1}{3}\end{cases}$};
    \node[below = of 11] (21) {\includegraphics[width=0.15\textwidth]{4F}}              node[above=-6mm of 21] {$\lambda^{(2)}$};
    \node[right = of 21] (22) {\includegraphics[width=0.15\textwidth]{2F2A1}}           node at ($(21)!0.5!(22)$) {$=$};
    \node[right = of 22] (23) {\includegraphics[width=0.15\textwidth]{2F2A2_A4_L1}}     node at ($(22)!0.5!(23)$) {$+$};
    \node[right = of 23] (24) {\includegraphics[width=0.15\textwidth]{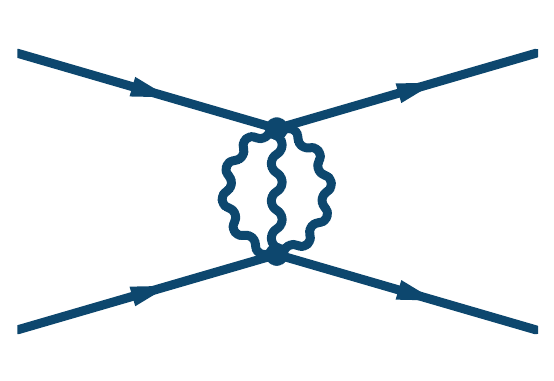}}     node at ($(23)!0.5!(24)$) {$+$};
    \node[right = of 24] (25) {\includegraphics[width=0.15\textwidth]{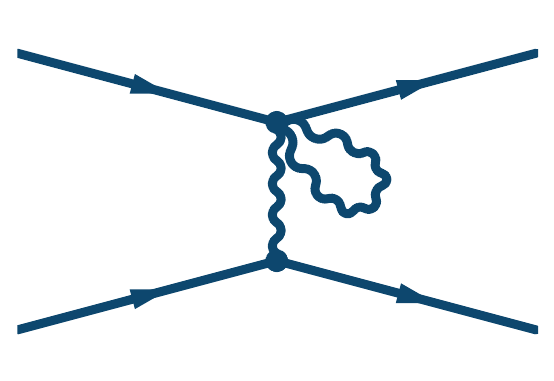}} node at ($(24)!0.5!(25)$) {$+$};
    \node[below=-5mm of 22] {\small $-c_1^2   \times \begin{cases}\frac{\gamma}{2}              \\\frac{1}{6}\end{cases}$};
    \node[below=-5mm of 23] {\small $-c_2^2   \times \begin{cases}\frac{1}{8}-\frac{\gamma^2}{2}\\\frac{2}{45}\end{cases}$};
    \node[below=-5mm of 24] {\small $-c_3^2   \times \begin{cases}\frac{3\gamma}{8}             \\\frac{1}{14}\end{cases}$};
    \node[below=-5mm of 25] {\small $-c_1 c_3 \times \begin{cases}\frac{1}{4}                   \\\frac{1}{5}\end{cases}$};
    \node[below =6mm of 21] (31) {\includegraphics[width=0.15\textwidth]{6F}}             node[above=-6mm of 31] {$\lambda^{(3)}$};
    \node[right = of 31] (32) {\includegraphics[width=0.15\textwidth]{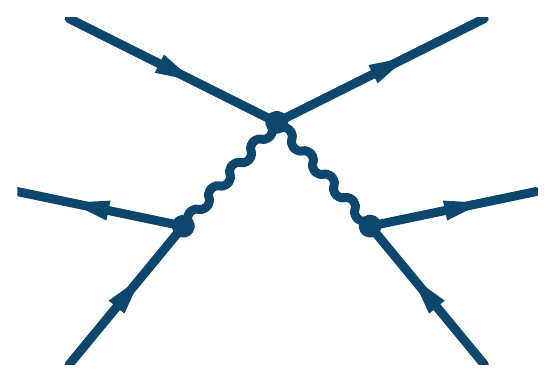}}     node at ($(31)!0.5!(32)$) {$=$};
    \node[right = of 32] (33) {\includegraphics[width=0.15\textwidth]{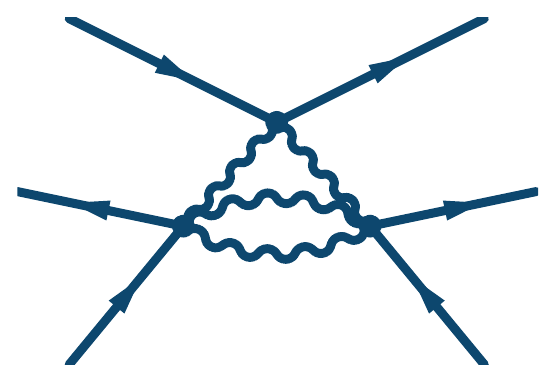}}     node at ($(32)!0.5!(33)$) {$+$};
    \node[right = of 33] (34) {\includegraphics[width=0.15\textwidth]{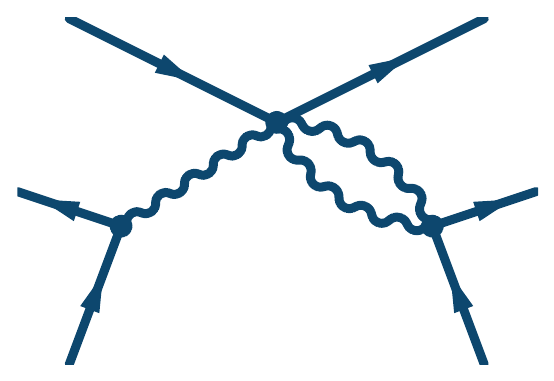}} node at ($(33)!0.5!(34)$) {$+$};
    \node[right = of 34] (35) {\includegraphics[width=0.15\textwidth]{3F2A2}}          node at ($(34)!0.5!(35)$) {$+$};
    \node[below=-3mm of 32] {\small $c_1 c_2^2   \times \begin{cases}\frac{1}{8}-\frac{\gamma^2}{2}    \\\frac{2}{45}\end{cases}$};
    \node[below=-3mm of 33] {\small $c_2 c_3^2   \times \begin{cases}\frac{5}{32}-\frac{3\gamma^2}{8}  \\\frac{2}{63}\end{cases}$};
    \node[below=-3mm of 34] {\small $c_1 c_2 c_3 \times \begin{cases}\frac{\gamma}{2}                  \\\frac{8}{105}\end{cases}$};
    \node[below=-3mm of 35] {\small $c_2^3       \times \begin{cases}\frac{\gamma^3}{3}                \\\frac{8}{2835}\end{cases}$};
  \end{tikzpicture}
  \caption{\label{fig:tbf-diagramatic}We present the results for the induced one-, two- and three-body forces for distributions with $N=2$ and $N = \infty$ with allowed auxiliary-fermion interactions presented in the upper right box ($c_{j>3}=0$).  The label below each diagram corresponds to the $c_j$-coefficient dependence of the induced effective forces $\lambda^{(k)}$.  The upper row of coefficients corresponds to $N=2$ and the lower row to $N=\infty$.  Here, we have defined $\gamma \equiv \Gamma(3/4) / \Gamma(1/4) \approx 0.34$.  Four-body and higher forces are not displayed and are, in general, non-zero.  Note that the wiggly lines correspond to fully dressed auxiliary field propagators.}
\end{figure}

The expansion of equation \eqref{eq:Zc} involves integration over the probability distribution times another polynomial with arbitrary powers of $\phi$
\begin{equation}
	\int\limits_{-\infty}^\infty \diff{\phi} e^{ - \phi^{2N}} \phi^{2k}
	=
	\frac{
		\Gamma\left(\frac{1 + 2 k}{2 N}\right)
	}{N}
	\overset{N\to\infty}{\longrightarrow}
	\frac{2}{1 + 2 k}
	\, , \quad
	\forall k \in \mathbb{N}_0, N \in \mathbb{N}
	\, .
\end{equation}
Note that odd powers in $\phi$ vanish by symmetry.
Using Fa\`{a} di Bruno's formula \cite{FdB1855}
\begin{multline}\label{eq:faadibruno}
	\frac{\partial^{k} \; }{\partial \phi^k} f_M(g_N(\phi))
	=
	k!
	\sum \limits_{\vec m \in \mathcal M^{(k)}}
	f_M^{(m_1 + \cdots + m_k)}(g_N(\phi))
	\prod\limits_{j=1}^k
	\left[
	\frac{1}{m_j!}
	\left(
		\frac{g^{(j)}_N(\phi)}{j!}
	\right)^{m_j}
	\right]
	\, , \\
	\mathcal M^{(k)}
	=
	\left\{
		\vec m \in \mathbbm{N}^k_0 \, \middle\vert \,
		\sum_{n=1}^k \, n \, m_n = k 
	\right\}
	\, ,
\end{multline}
we simultaneously expand equation \eqref{eq:Zc} in $\phi$ and $\hat \rho$ and equation \eqref{eq:Zla} in $\hat \rho$ only
\begin{align}\label{eq:Zcresfull}
	\Z_{c, N}^{(M)}
	&=
	\sum\limits_{k = \left \lceil{M/2}\right \rceil }^{\left \lfloor{(2N-1)M/2}\right \rfloor}
	\frac{\Gamma\left(\frac{1 + 2k}{2 N}\right)}{ \Gamma \left( \frac{1}{2N}  \right) }
	\sum \limits_{\vec m \in \M_{NM}^{(2k)}}
	\prod\limits_{j=1}^{2k}
		\left[
			\frac{ (-c_j)^{m_j} }{m_j!}
		\right]
	\, , & 
	\Z^{(M)}_{\lambda}
	&=
	\sum \limits_{\vec m \in \M^{(M)}}
	\prod\limits_{k=1}^M 
	\left[
		\frac{ \left(- \lambda^{(k)} \right)^{m_k} }{m_k!} 
	\right]
	\, .
\end{align}
Here we have defined the set
\begin{align}
	\M_{NM}^{(2k)}
	=&
	\mathcal M^{(2k)} \cap
	\left\{
		\vec m \in \mathbbm{N}^{2k}_0 \, \middle\vert \,
		\| \vec m \| = M
		\land\left(j\geq 2N \Rightarrow m_j=0\right)
	\right\}
	\, ,
\end{align}
where $\| \vec m \|$ denotes the sum over all components of $\vec m$.
The set $ \M^{(M)} $ is the same as defined in equation \eqref{eq:faadibruno}.
Because one can easily verify\footnote{In order to obtain the coefficient $\lambda^{(M)}$, one needs that $m_M \geq 1$. But because $\vec m \in \M^{(M)}$, we have that $1 \, m_1 + 2 \, m_2 + \cdots + M \,  m_M = M$ which means $m_M = 1$ and $ m_j = 0$ $\forall j\neq M$.} that the $\Z^{(M)}_{\lambda}$ is linear in $\lambda^{(M)}$, one can recursively determine the dependence of $\lambda^{(M)}$ on the previous coefficients $\lambda^{(k)}$ with $k < M$ and the expansion in the $c$ coefficients by requiring
\begin{align}
	\Z^{(M)}_{\lambda}
	&=
	-\lambda^{(M)} 
	+
	\Z^{(M)}_{\lambda}\big|_{\lambda^{(M)}\to 0}
	\overset{!}{=}
	\Z^{(M)}_{c,N}
	\, ,
	&
	\Z^{(1)}_{\lambda}\big|_{\lambda^{(1)}\to 0}
	&=
	0
	\, .
\end{align}
We want to highlight that all of the obtained results are smooth in the limit of $N \to \infty$, which would allow one to incorporate an infinite amount of freely tuneable $c_j$-coefficients.
In this limit, the probability distribution for the auxiliary fields becomes uniform: $P_\infty(\phi) = 1/2$ for $\phi \in (-1, 1)$ and zero otherwise.
However, for every different order $N$, the dependence of the induced forces $\lambda^{(k)}$ on the auxiliary coefficients $c_j$ slightly changes.
We present an example with $N=2$ and $N = \infty$ in figure \ref{fig:tbf-diagramatic}.

Because the analytic form of the induced forces that depend on the auxiliary field interaction parameters can be tuned in a completely general fashion, we provide a {\it Mathematica} notebook in the Supplementary Material of Ref.~\cite{Korber:2017emn} which can be used to compute the $\lambda^{(M)}$ for the problem of choice.

Furthermore, we emphasize that the infinite tower of induced interactions can be systematically controlled.
It can be inductively shown\footnote{A probably more intuitive argument can be directly read off figure \ref{fig:tbf-diagramatic}: because all allowed auxiliary field-fermion interactions are proportional to the fermion bilinear, an $N$-body fermion interaction is exactly proportional to $N$ auxiliary field-fermion interactions and thus must contain $N$ $c_j$ coefficients.} that the coefficient $\lambda^{(k)} \propto c_{i_1} c_{i_2} \cdots c_{i_k}$.
Thus, if the magnitude of $c_j$ coefficients is smaller than one, higher forces become increasingly smaller.

\section{Numerical Results}\label{sec:numerics}
In order to demonstrate the applicability of the proposed method for including $N$-body forces in lattice algorithms, we set up a toy problem which can be solved analytically and compare numerically sampled results to their analytical counterparts.
As a test system, we choose a two-site model (sites $i=0$ and $1$) and place up to 3 different fermion species $f$ at each site.
The Hamiltonian which describes this system is given by
\begin{equation}\label{eqn:hamiltonian}
\hat H=\kappa\left(-3\sum_{f=1}^3\sum_{\langle i,j\rangle}\hat a_{f,i}\hat a^\dag_{f,j}+\lambda^{(2)}\sum_{i\in\{0,1\}} \hat \rho_i^2+\lambda^{(3)}\sum_{i\in\{0,1\}} \hat \rho_i^3\right)\ .
\end{equation}
Here, $\hat a_{f,i}$ ($\hat a_{f,i}^\dagger$) annihilates (creates) a fermion of species $f$ at site $i$ and $\hat \rho_i=\sum_f \hat a^\dag_{f,i}\hat a_{f,i}$ is the fermion density or number operator at site $i$.
The kinetic part of the Hamiltonian sums over nearest neighbours $\langle i,j\rangle \in \{ (0,1), (1,0) \}$, while the strength of the two-body force $\lambda^{(2)}$ and three-body force $\lambda^{(3)}$ will take different values in order to explore the capabilities of the proposed transformation.
We analytically compute the largest eigenvalue of the transfer matrix $\hat T(\tau) \equiv \, : \exp ( - \tau \hat H ):$ (see \cite{Lee:2008fa}), and compare the result to the large time projection of the transfer matrix acting on an initial target state $\Psi_T$
\begin{multline}\label{eqn:E}
	E
	=
	-\lim_{\tau\to\infty}{\partial_\tau \log Z[\tau,\Psi_T]} \, ,
	\\
	Z[\tau,\Psi_T] 
	\equiv
	\langle \Psi_T| : e^{-\tau \hat H} :|\Psi_T\rangle 
	= 
	\left(\prod_{x} \int d\phi_x P_N(\phi_x)\right) \K[\tau,\vec{\phi},c_j,\Psi_T]
	\, .
\end{multline}
We compute the large time projected matrix element using the proposed prescription for auxiliary fields.
In other words we express the Hamiltonian of equation \eqref{eqn:hamiltonian} in terms of effective couplings $c_j$ and fermion-bilinear auxiliary field interactions\footnote{Note that we have subtracted the effects of the  induced one-body force directly on the level of auxiliary fields.  Already the agreement of one-body energy levels in this framework is a non-trivial cross-check.}.
The coefficients $c_j$ are matched to the effective two- and three-body forces by the relations listed in figure \ref{fig:tbf-diagramatic} and the remaining integral is computed by sampling the auxiliary field according to distribution $P_N(\phi)$ (same distribution for both lattice sites).
Finally the function $\K$ is obtained by sandwiching the auxiliary-field-transformed transfer matrix with an (arbitrary) initial wave function, which we have chosen as $\ket{\Psi_T}=\prod_{f=1}^{F}\frac{1}{\sqrt{2}}[a^\dag_{f,0}-a^\dag_{f,1}]\ket{0}$.

\begin{figure}[t!]
\includegraphics[width=\textwidth]{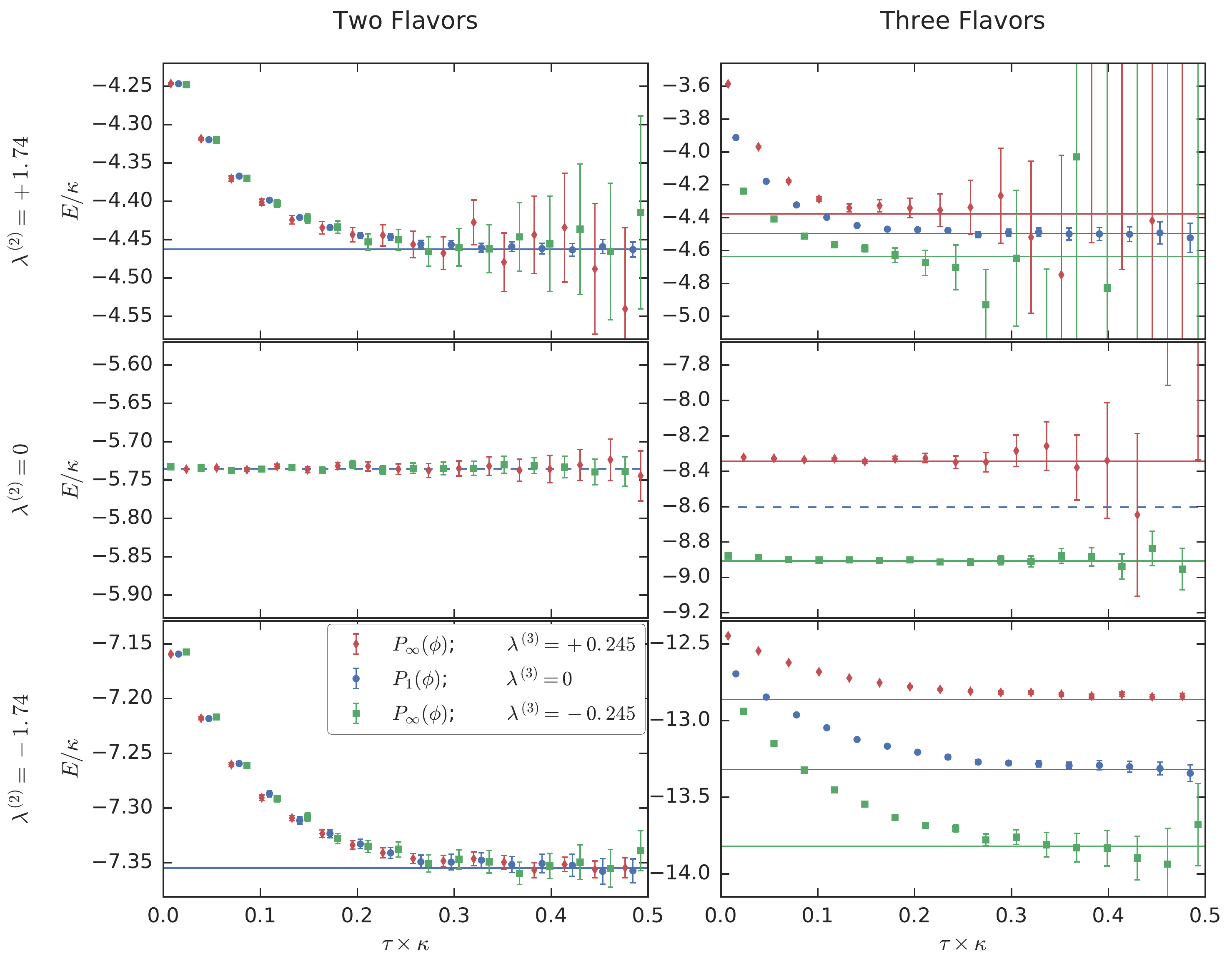}
\caption{$E(\tau)/\kappa$ for different systems.    All systems were studied with $2\times10^8$ measurements.  Some systems show a
  signal-to-noise problem and others yield an easy extraction of a
  constant plateau in the long-time-limit.  The left panels all show
  two-fermion systems and the corresponding right panels show the
  three-fermion systems with the same parameters.  In the top (middle)
  [bottom] two panels we show a system with repulsive (absent)
  [attractive] two-body forces.  Red (blue)
  [green] points correspond to repulsive (absent) [attractive] three
  body forces.  The blue points were sampled according to the HS
  distribution $P_1$ and the other points according to $P_\infty$.
  The data in the middle two panels were generated with $P_\infty$ and coefficients
  $c_j = 0 $ for $j > 3$ tuned to exactly cancel the two-body force, and
  we show dashed lines for the corresponding non-interacting energies.
  This figure is taken from \cite{Korber:2017emn}.
 }
\label{fig:spectrum}
\end{figure}

The comparison between analytical and stochastic computations is demonstrated for two- and three-fermion systems $F=2$ and $3$, for the regular Hubbard-Stratonovich transformation ($N=1$) and generalized transformation following the uniform distribution ($N=\infty$), for attractive, absent and repulsive two-body forces $\lambda^{(2)} \in \{ -1.74, 0.0, +1.74 \}$ in case of both transformations and attractive, absent and repulsive three-body forces $\lambda^{(3)} \in \{ -0.245, 0.0, +0.245 \}$ in case of the generalized transformation with $N = \infty$.
Because we are free to pick any values for the infinite number of coefficients $c_j$ in case of the $N=\infty$ transformation, we arbitrarily pick $c_j = 0$ for $j>3$.
The results of the comparison can be found in figure \ref{fig:spectrum}.
Solid lines correspond the the analytical results and data points to the sampled auxiliary field computations.
Error bars correspond to the stochastic uncertainty of sampled observables.
Data points at a given time step $\tau$ for different values of the three-body force are slightly shifted for visualisation purposes.

All of the results, in the large-time limit, agree well with the analytical results within uncertainties.
However, computations for specific values of the induced two- and three-body forces suffer from relatively large statistical uncertainties.
This is expected and can be explained by the Hamburger truncated momentum problem \cite{adamyan2003reconstruction, Chen:2004rq}.
In forthcoming work \cite{our-long-paper} we intend to analyse such fluctuations in terms of signal-to-noise behaviour and their relation to the sign problem in more detail.

\section{Discussion}\label{sec:discussion}
We have presented a general prescription which generalizes the Hubbard-Stratonovich transformation such that any $N$-body contact interactions for fermions (and similarly for bosons) can be cast to a set of linearized fermion (boson) interactions with an auxiliary field.
Depending on the distribution of auxiliary fields, ranging from gaussian to uniform, more possible interactions with the auxiliary field can be included.
Each of the linear interactions with the auxiliary field comes with an a priori freely tunable interaction coefficient that can be analytically determined to reproduce the original forces in a controled fashion after integrating out the auxiliary fields.
If one chooses a proper sampling distribution and as a consequence obtains a sufficient amount of auxiliary interactions, the set of different interaction coefficients which reproduce the original forces is infinite.

We verify these predictions for a simple two-site model and confirm that statistical simulations, using the presented prescription, converge against their analytical result within uncertainties.
This non-trivial cross-check is demonstrated for a set of attractive, absent and repulsive two- and three-body interactions.
We find that certain combinations of interaction coefficients are more favourable than others in terms of convergence of statistical observables: e.g. one should prefer real over complex coefficients.
Furthermore, for certain effective forces, it is not possible to find sets of interaction coefficients which do not suffer from large statistical fluctuations.

For upcoming work, we intend to further investigate the signal-to-noise behaviour for different sets of interaction coefficients.
Also, it might be interesting to generalize this transformation even further to include non-central and non-local interactions or understand the renormalisation induced by the effective forces on the level of interaction coefficients.

Last but not least, we expect that our findings impact a variety of physical systems for which the inclusion of many-body forces is relevant.
Examples include the non-perturbative addition of three-body forces in nuclear lattice effective field theory, the study of systems near the Efimov threshold and studies of halo nuclei.

\bibliography{references}

\end{document}